\documentclass[doublecol,linenumbers]{epl2}  
\usepackage{epsfig}
\usepackage{amsmath}

\begin{document}

\title{$\Psi(2S)$ production in $p+A$ collisions}

\author{Partha Pratim Bhaduri\inst{1}
\and 
Abhijit Bhattacharyya\inst{2}
        }

\institute {
\inst{1} {Variable Energy Cyclotron Centre, HBNI, 1/AF Bidhan Nagar, Kolkata 700064, India}  

\inst{2} {Department of Physics, University of Calcutta, 92, A. P. C. Road, Kolkata - 700009, India}
           
}


\abstract
{
We have analyzed the data of $\psi(2S)$ production in proton-nucleus ($p+A$) collisions, available from the NA50 Collaboration in the SPS energy domain. The investigated data sets include the absolute production cross sections as well as $\psi(2S)$-to-Drell Yan (DY) cross section ratios. An adapted version of two component QVZ model has been employed to calculate $\psi(2S)$ production cross sections. For both $\psi(2S)$ and DY production, nuclear modifications to the free nucleon structure functions are taken into account.  
For $\psi(2S)$, final state interaction of the produced $c\bar{c}$ pairs with the nuclear medium is also taken into account, in accordance with the previously analyzed $J/\psi$ data. A satisfactory description of the data in $p+A$ collisions is obtained.  Model calculations are extrapolated to predict the $\psi(2S)$ suppression in proton induced collisions at near threshold energies.
}

\maketitle
\section{Introduction}

Charmonium suppression in nuclear collisions is one of the oldest signatures to indicate the formation of quark-gluon plasma (QGP)~\cite{MS}. However, to identify the genuine plasma effects, a precise estimation of the suppression induced by the cold nuclear medium (CNM) is a necessary prerequisite~\cite{Vogt,Satz,Andronic:2014sga}. Over the years charmonium production has thus been studied at different fixed target and collider facilities, in proton-nucleus ($p+A$) collisions, where formation of a deconfined medium is usually not expected. A precise understanding of this so called ``normal'' suppression is crucial to establish a robust baseline, with respect to which one can isolate the ``anomalous'' suppression pattern, specific to the dense QCD medium produced in heavy-ion collisions.

Among different charmonium states $J/\psi$ is most extensively studied in nuclear collisions. However not only $J/\psi$, but other charmonium states (e.g. $\psi(2S)$, $\chi_{c}$) are also suppressed in a QGP or in a nuclear matter. In our earlier work~\cite{partha1}, we have predicted the $J/\psi$ suppression in $p+A$ collisions at Facility for Aniti-proton and Ion Research (FAIR) energy domain, within the ambit of modified QVZ model~\cite{Qui}. Different parameters of the  model were fixed using the available data on $J/\psi$ production in $p+p$ and $p+A$ collisions from different fixed target experiments. Such calibrated model was then extrapolated to the FAIR energy domain. In the present article, we plan to check the viability of our model by examining the available data on $\psi(2S)$ production in $p+A$ collisions at SPS~\cite{NA50-400,NA50-450}. After successful description of the SPS data, calculations are extrapolated to FAIR SIS100 energy regime, to estimate the CNM suppression in $\psi(2S)$ production. 

\section{A brief description of the model}

\begin{figure*} 
\includegraphics[height=5.5cm,width=5.5cm]{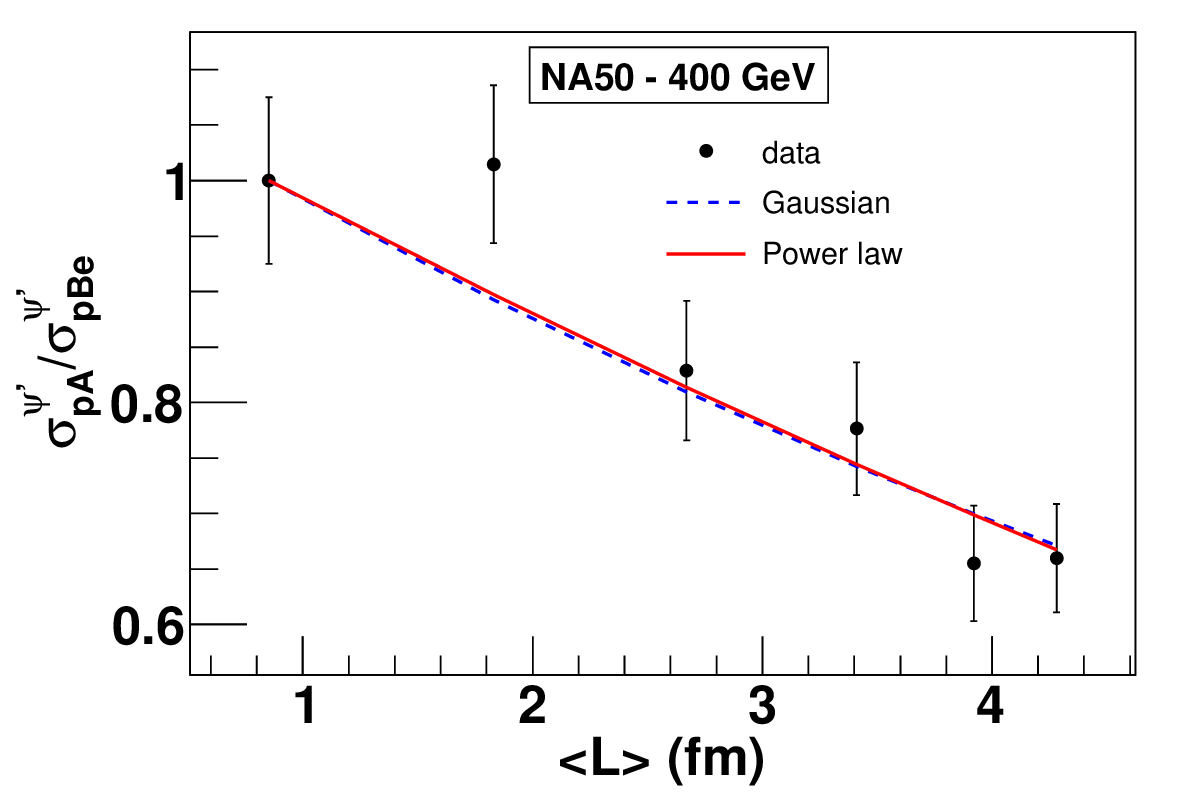}
\includegraphics[height=5.5cm,width=5.5cm]{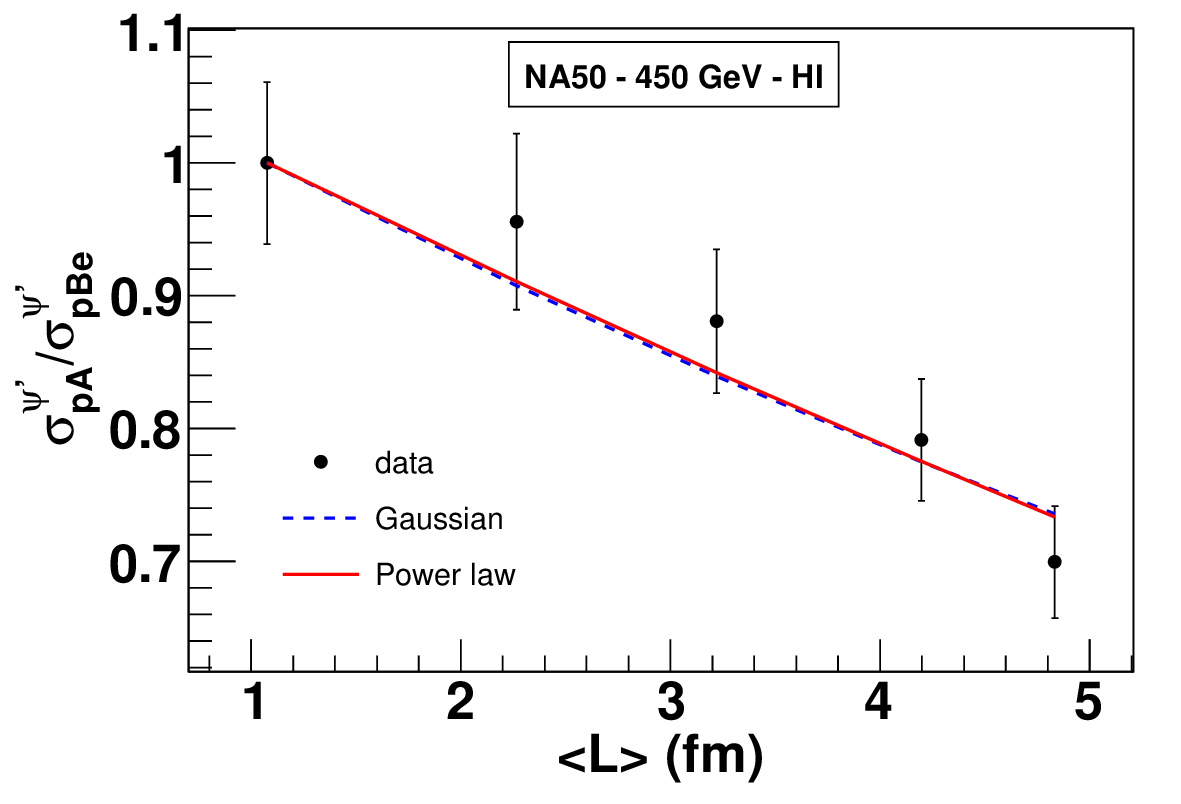}
\includegraphics[height=5.5cm,width=5.5cm]{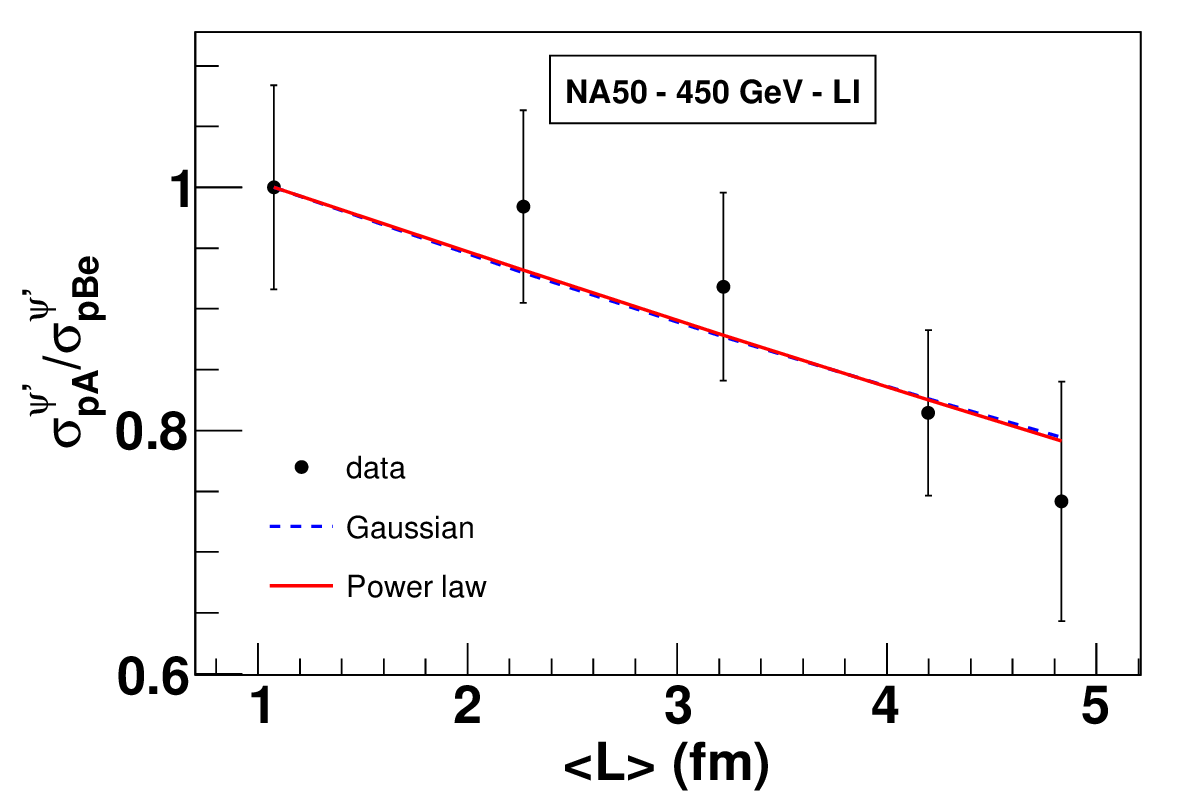}
\caption{\footnotesize (Color online) Model description of $\psi(2S)$ production in $p+A$ collisions at 400 and 450 GeV incident energy of the proton beams. Data are represented as the ratio of $\psi(2S)$ production cross sections in $p+A$ collisions to that in $p+Be$ collisions. As usual practice, for 450 GeV beam energy, data for high intensity (HI) and and low intensity (LI) runs are shown separately. At 450 GeV data were collected for five different target nuclei ($Be, Al, Cu, Ag, W$), while six different target nuclei ($Be, Al, Cu, Ag, W, Pb$) were used for 400 GeV proton beam. The two theoretical curves represent two different parametric forms of $\psi(2S)$ formation probability ($F(q^{2})$).}
\label{fig1}
\end{figure*}

For calculation of $\psi(2S)$ production cross sections, we have used the adapted version of the originally proposed QVZ model~\cite{Qui}. Details of the original model and its modifications as used in the present calculations can be found in refs.~\cite{partha1,partha2,Qui,ch02,ch02b}. Here we present a brief description for the sake of completeness. In hadronic collisions, $\psi(2S)$ production is assumed to be factorised into two steps. The initial stage is the production of the $c\bar{c}$ pair described by leading order (LO) perturbative QCD. At leading order the dominant contribution to $c\bar{c}$ production comes from $q\bar{q}$ annihilation and $gg$ fusion. This is followed by the non-perturbative formation of the physical resonance in the second stage. Resonance formation probabilities are incorporated in QVZ model using different functional forms which account for a wide variety of color neutralisation mechanisms. The single differential $\psi(2S)$ production cross section in collisions of hadrons  $h_1$ and $h_2$, at the centre of mass energy $\sqrt{s}$ reads as,

\begin{equation}
\label{diff}
\frac{d\sigma_{h_1h_2}^{\psi(2S)}}{dy_{cms}} = K_{\psi(2S)}\int dQ^{2}\left(\frac{d\sigma_{h_1h_2}^{c\bar{c}}}{dQ^2dy_{cms}}\right)\times F_{c\bar{c}
\rightarrow \psi(2S)}(q^2)
\end{equation}
\noindent  where $Q^2 = q^2 +4 m_C^2$ with $m_C = 1.5$ GeV being the mass of the charm quark and $y_{cms}$ is the centre of mass rapidity of the $c\bar{c}$ pair. The 
term $K_{\psi(2S)}$ accounts for effective higher order contributions and $F_{c\bar{c} \rightarrow  \psi(2S)}(q^2)$ is the transition probability of a $c\bar{c}$ pair with relative momentum square $q^2$ to evolve into a physical $\psi(2S)$ meson. Two different functional forms namely Gaussian form ($F^{\rm (G)}(q^2)$) and power law form ($F^{\rm (P)}(q^2)$) mimicking two different mechanisms of colour neutralisation, were found earlier to describe the absolute $J/\psi$ production cross section data in $p+A$ collisions reasonably well~\cite{partha1}. Here we take the same two forms of $F(q^2)$, for calculating $\psi(2S)$ production cross section. They read as:
\begin{eqnarray}
F^{\rm (G)}_{c\bar{c}\rightarrow{\psi(2S)}(q^2)}
&=& N_{\psi(2S)}\, \theta(q^2)\,
\exp\left[-q^2/(2\alpha^2_F)\right]\ ,
\label{gauss}
\\
F^{\rm (P)}_{c\bar{c}\rightarrow{\psi(2S)}(q^2)}
&=& N_{\psi(2S)}\, \theta(q^2)\, \theta(4m_D^2-4m_C^2-q^2)
\nonumber \\
&\times & \left(1-q^2/(4m_D^2-4m_C^2)\right)^{\alpha_F}\ ,
\label{power}
\end{eqnarray}
where $m_D = 1.85$ GeV is the mass scale for the open charm production threshold and $N_{\psi(2S)}$ and $\alpha_F$ are two parameters of $F_{c\bar{c} \rightarrow  \psi(2S)}(q^2)$, which should of course be different compared to those of $J/\psi$. They are fixed by comparing the model results with the available data. Introduction of $f_{\psi(2S)}$ defined as $f_{\psi(2S)} = K_{\psi(2S)} \times N_{\psi(2S)}$ helps in reducing the number of free parameters. We should mention here that in formulating the transition probability we 
have not considered the node in the $\Psi(2S)$ wave function and used the same forms for both $J/\psi$ and $\Psi(2S)$~\cite{hufner}.
\begin{figure*} 
\includegraphics[height=5.5cm,width=5.5cm]{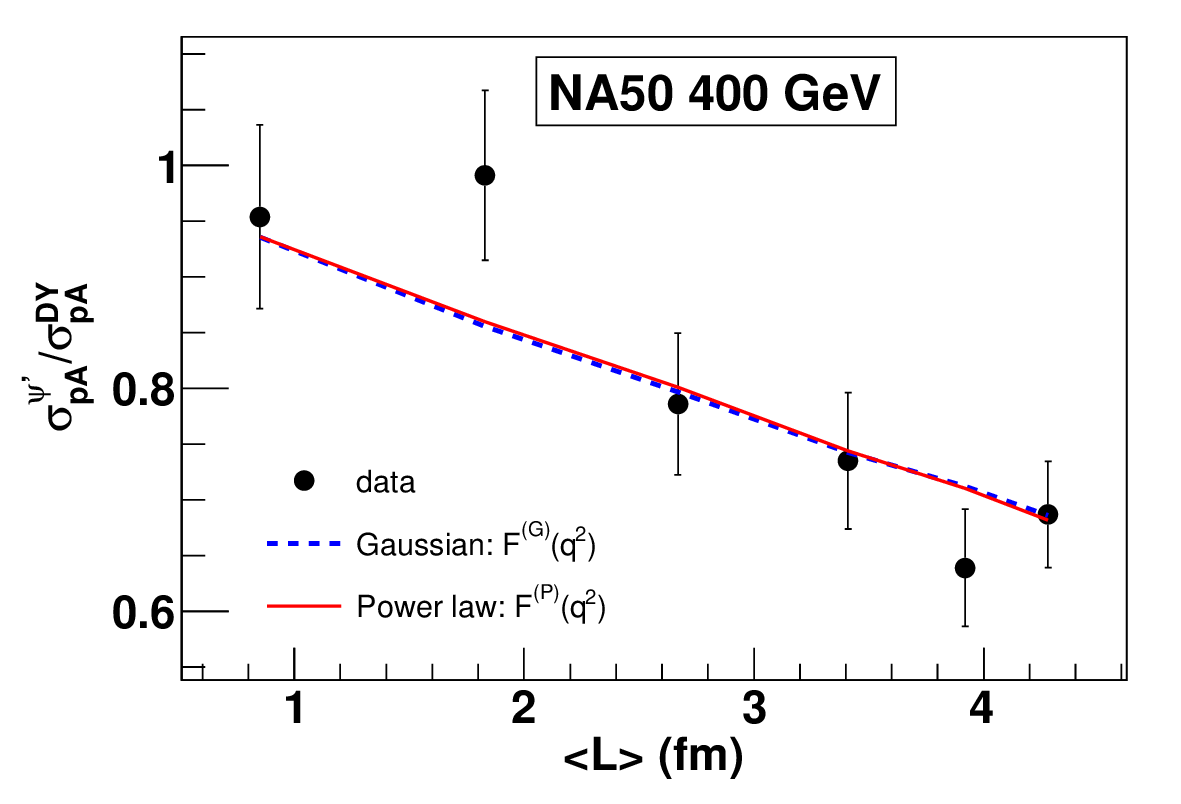}
\includegraphics[height=5.5cm,width=5.5cm]{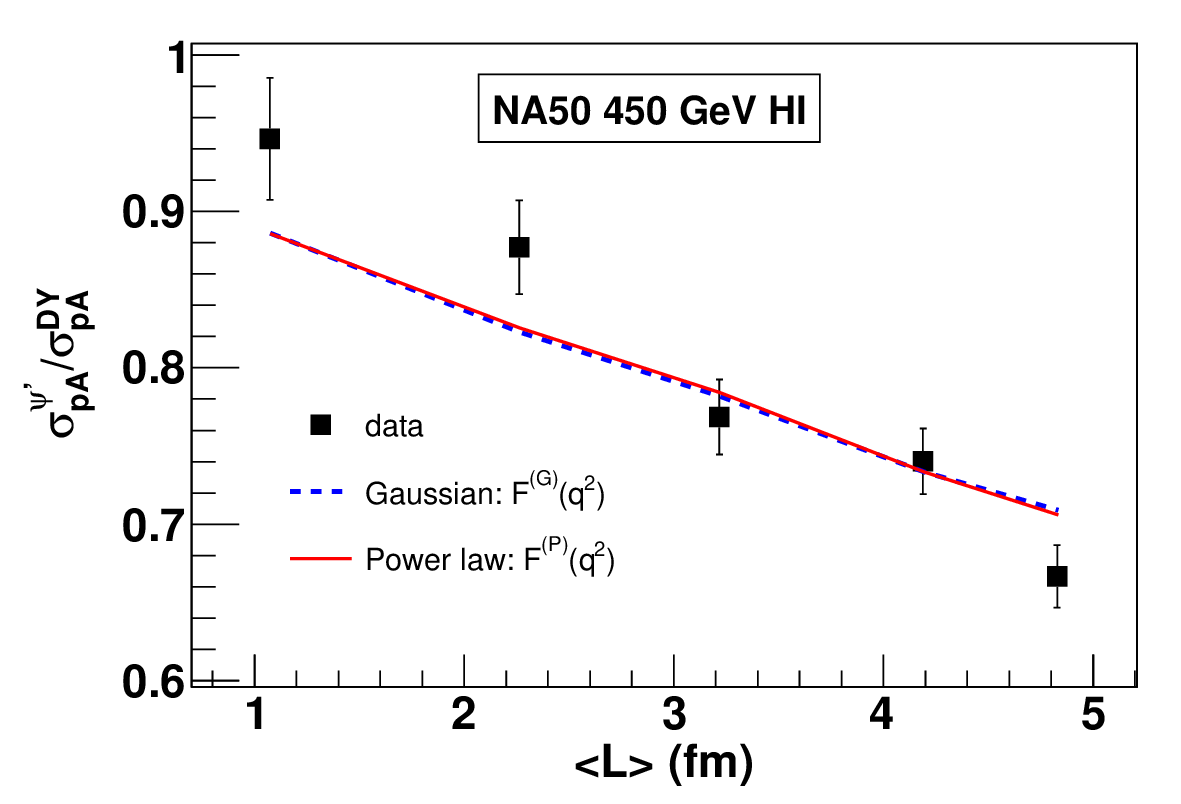}
\includegraphics[height=5.5cm,width=5.5cm]{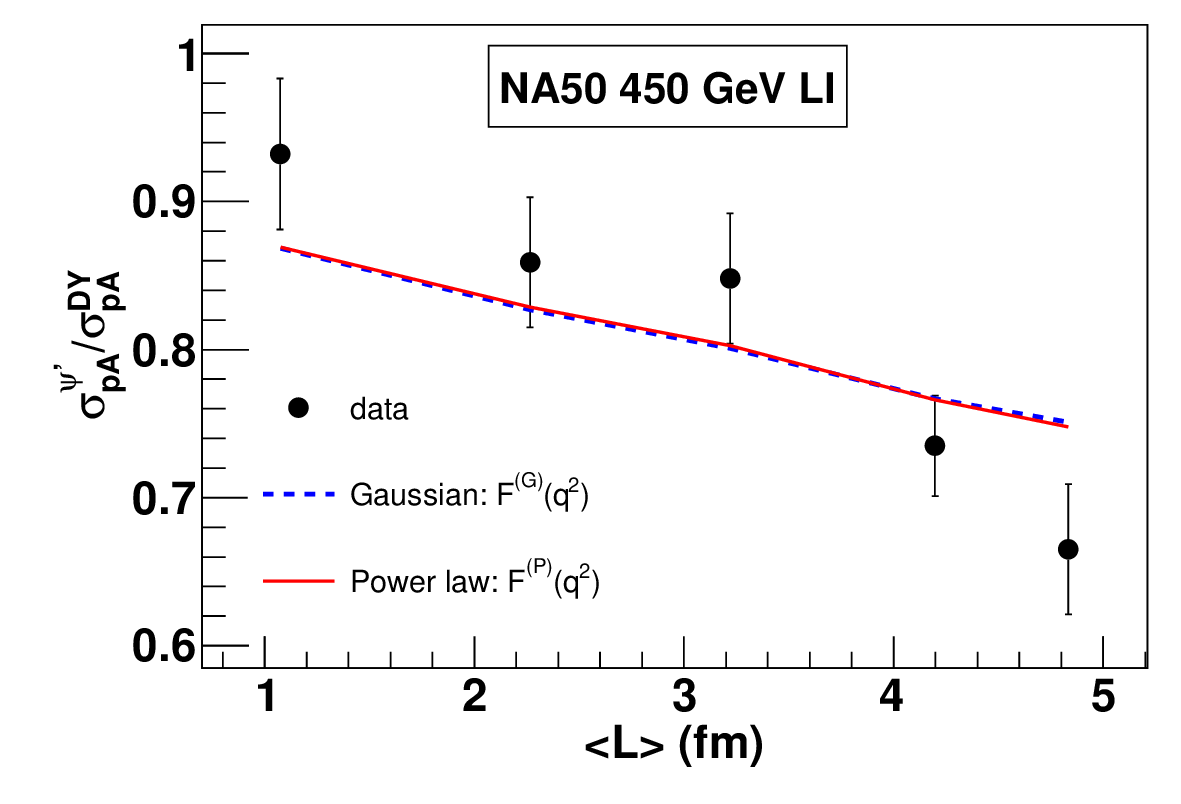}
\caption{\footnotesize (Color online) Model description of the data on ratio of the $\psi(2S)$-to-Drell-Yan production cross-sections in 450 GeV and 400 GeV $p+A$ collisions, as published by NA50 Collaboration at SPS.} 
\label{fig2}
\end{figure*}

In $p+A$ collisions, $\psi(2S)$ production cross sections are modified because of cold nuclear matter effects. At the initial stage, nuclear modifications of the parton densities inside the target modify the $c\bar{c}$ pair production cross section. In our analysis, we opted for leading order (LO) MSTW2008~\cite{MSTW} central set with minimum uncertainties, for free proton parton distribution function (PDF) and LO EPS09~\cite{EPS09} interface to account for the nuclear effects.  
A detailed account of these initial state modifications of the parton distributions as they are implemented in our model calculations can be found in ref.~\cite{partha1}. 

The nascent $c\bar{c}$ pairs produced via partonic hard scattering experience multiple soft collisions with nuclear medium during their passage through the target. As a result the 
pairs gain in energy and hence in invariant mass. In this process, some of the $c\bar{c}$ pairs can gain enough mass to cross the threshold to transmute to two open charm mesons. 
This results into the reduction in $\psi(2S)$ yield compared to the nucleon-nucleon collisions. The overall effect of the multiple scattering of the $c\bar{c}$ pairs inside the target is 
represented by a shift of $q^2$ in the transition probability~\cite{Qui,BQP}, 

\begin{equation}
q^2  \longrightarrow  \bar{q}^2 = q^2 + \varepsilon^2\, <L(A)> \ .
\label{q2shift}
\end{equation}
where $\epsilon^2$ is the increase in the square of the relative four momentum of an evolving $c\bar{c}$ pair per unit path length, inside the nucleus and $<L(A)>$ is the mean geometrical path length traversed by the $c\bar{c}$ pair inside the nuclear medium, from its point of production until it exits the medium.

Incorporation of final state dissociation in QVZ framework is largely different from the conventional approach. In usual practice, the nuclear dissociation of the different charmonium 
states are treated within Glauber model with an absorption cross section $\sigma_{\rm abs}$ quantifying the amount of dissociation. Note that, in the present approach, the nuclear effects are operative on the pre-resonant $c\bar{c}$ pairs which are yet to be hadronized. Hence for a given kinematic domain, the value of $\epsilon^2$ should be same for both $J/\psi$ and $\psi(2S)$ formation. We have seen in ref.~\cite{partha1} that for both parameterisations of $F(q^2)$, the corresponding values of $\epsilon^2$ exhibited non-trivial dependence on the beam energy ($E_b$).  Lower the beam energy, smaller would be the velocity of the colliding nuclei leading to larger collision time. Hence the CNM effects would be operative over a longer time and the charmonia during its evolution is more likely to encounter the nuclear medium.  So as the  beam energy is lowered, the value of $\epsilon^2$ increases, implying larger nuclear dissociations. This feature is also in-line with existing theoretical and experimental observations which (within a conventional Glauber model framework) report a larger J/$\psi$ absorption cross section at lower beam energies~\cite{Lourenco,NA60}. For a more quantitative  information on values of $\epsilon^2$ at different beam energies and for different parameterizations, see Table III and Fig.13 of ref.~\cite{partha1}.

\section{Analysis of SPS data}
\begin{figure*} 
\includegraphics[height=5.5cm,width=5.5cm]{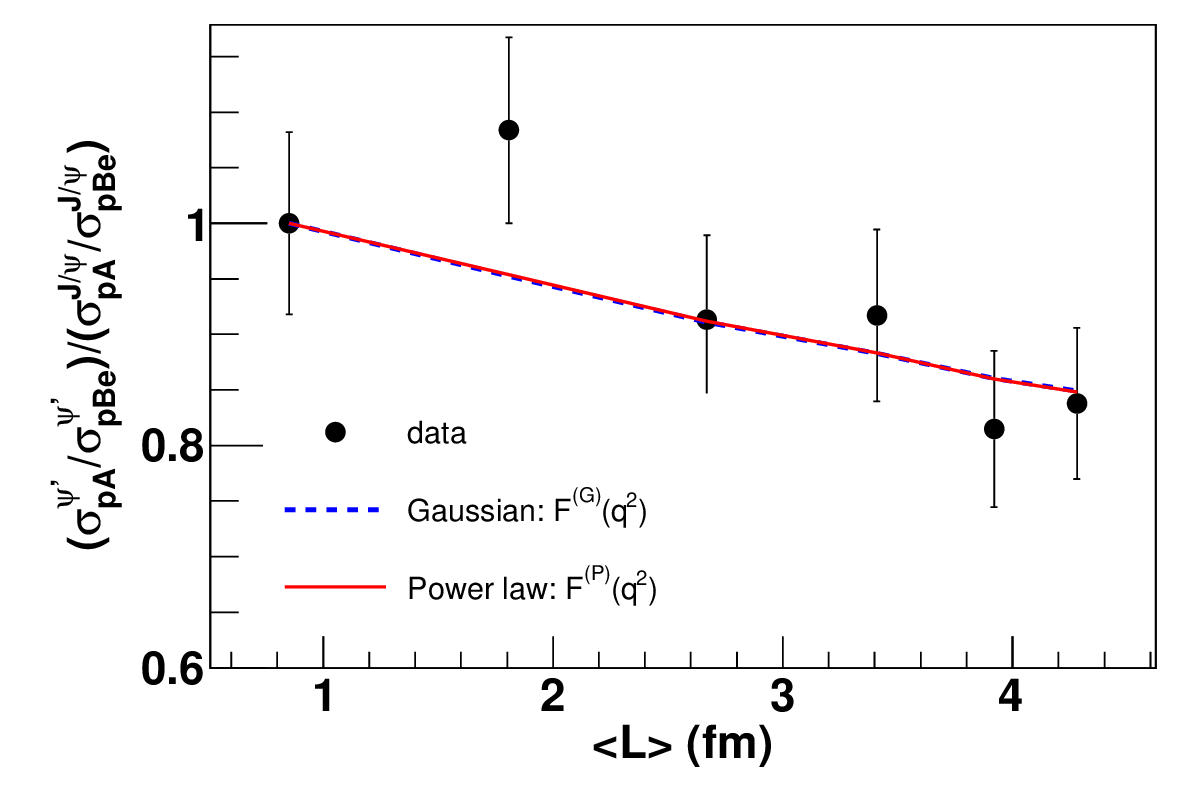}
\includegraphics[height=5.5cm,width=5.5cm]{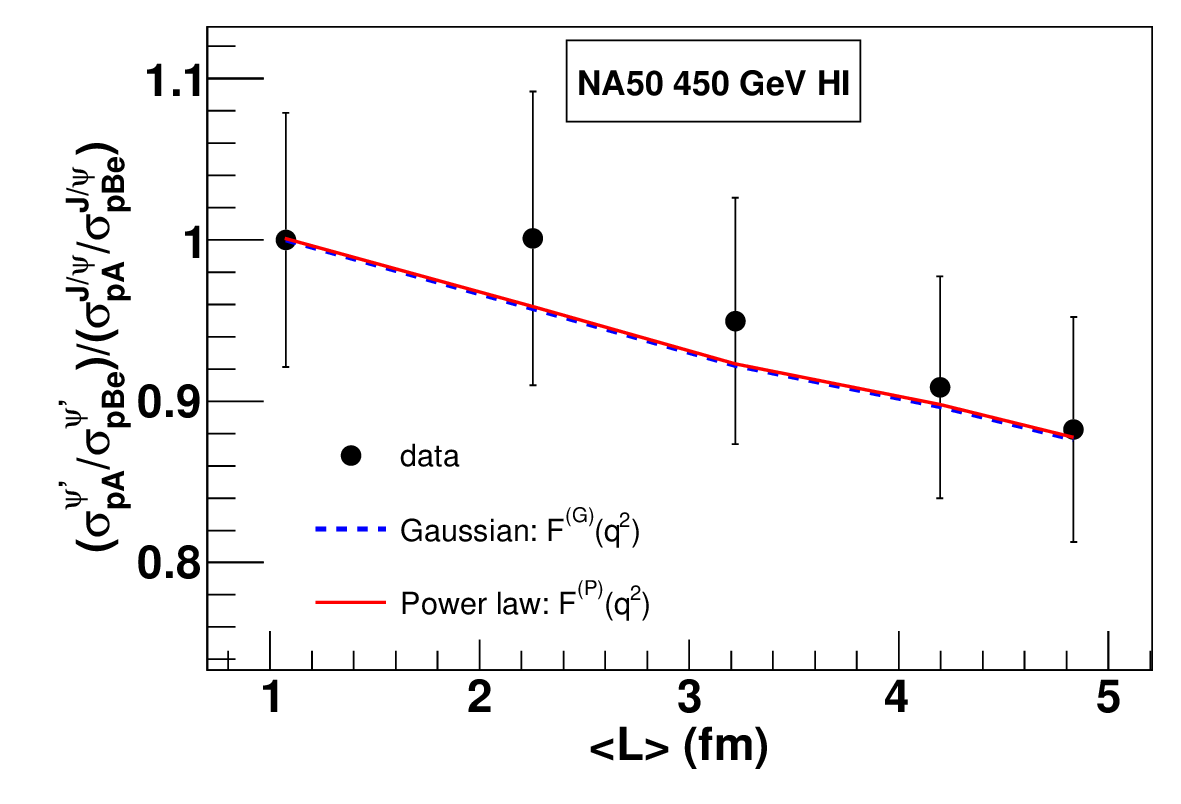}
\includegraphics[height=5.5cm,width=5.5cm]{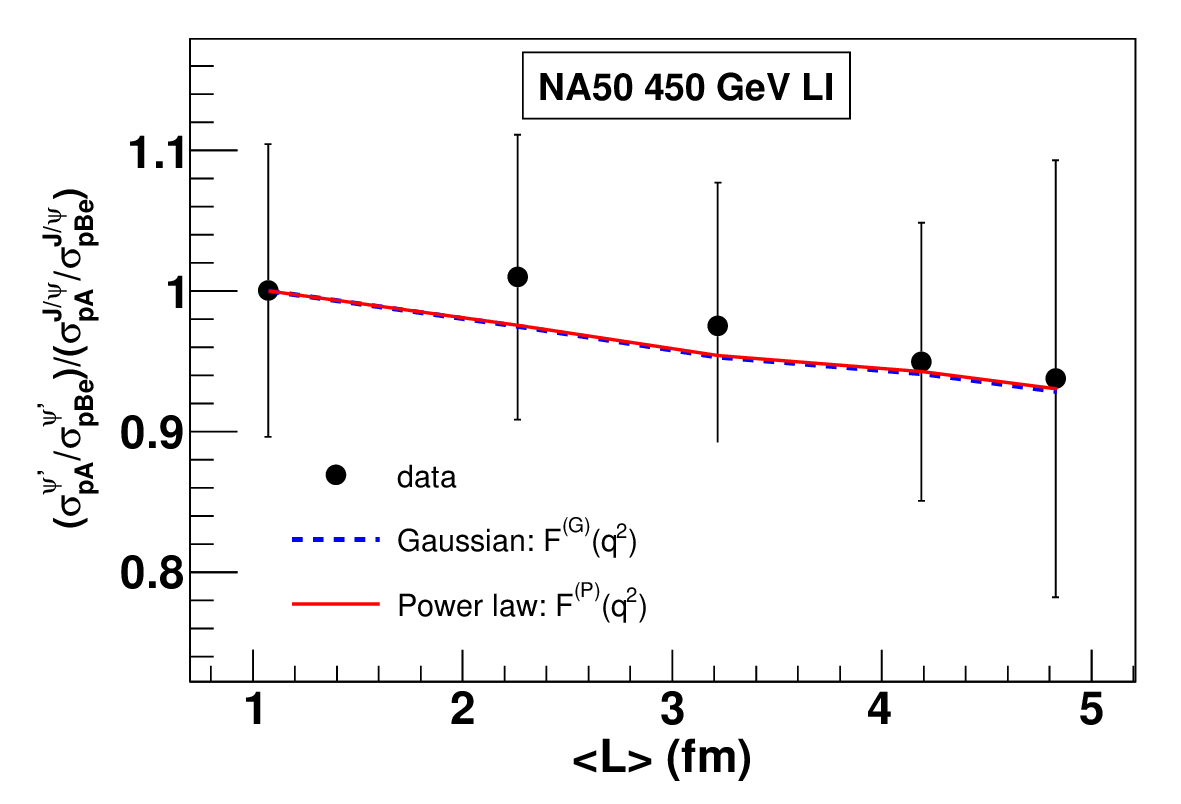}
\caption{\footnotesize (Color online) Model description of the data on the double ratio of the $\psi(2S)$-to-$J/\psi$ production cross-sections in 450 GeV and 400 GeV $p+A$ collisions. The double ratios are derived from the measured $J/\psi$ and $\psi(2S)$ cross sections in $p+A$ collisions.}
\label{fig3}
\end{figure*}

 With this brief description of the model, we now move forward to examine the viability of the model calculations in describing the available data on $\psi(2S)$ production in low energy $p+A$ collisions. Because of the much smaller production cross sections, measured $\psi(2S)$ yields are not as abundant as $J/\psi$ production, particularly at low energy fixed target experiments. At SPS a systematic measurement is available only from NA50 Collaboration, which measured $\psi(2S)$ production via their di-muon decay channel in 450 GeV~\cite{NA50-450} and 400 GeV~\cite{NA50-400} $p+A$ collisions. With 450 GeV proton beam, NA50 collected two sets of $p+A$ data in two independent runs with different beam intensities. The first set was collected with a high intensity (HI) 450 GeV proton beam using five different targets ($Be, Al, Cu, Ag, W$). Subsequently at the same beam energy new $p+A$ data samples were collected with same set of targets with low beam intensity (LI) and having 20 - 30$\%$ of the statistics of the HI set. Though they were initially collected in a slightly different kinematic domain, for a coherent comparison all the data sets at 450 GeV are corrected for a common phase space window: $-0.5 < y_{cms} <0.5$ and $-0.5 < cos(\theta_{cs}) < 0.5$, where $y_{cms}$ denotes di-muon centre-of-mass rapidity and $\theta_{cs}$ is the Collins-Soper angle. At 400 GeV, data were collected for six different targets ($Be, Al, Cu, Ag, W, Pb$) in the kinematic domain: $-0.425 < y_{cms} < 0.575 $ and $-0.5 < cos(\theta_{cs}) < 0.5$. Even though the beam intensities were slightly higher than that of 450 GeV HI data samples, 
short beam time prevented the collection of sizeable statistics. In 400 GeV run, unlike previous measurements, data for all targets were collected in the same data taking period leading to significant reduction in the systematic errors. Later NA60 Collaboration studied charmonium production in $p+A$ collisions with proton beams of 400 and 158 GeV. Due to limitation in statistics their measurements were limited to $J/\psi$ production only.

Here a few words on the limitations of the analysis of $\psi(2S)$ production in low energy collisions may be required. For describing the production in $p+A$ collisions, the QVZ model has effectively three parameters namely $f_{\psi(2S)}$ and $\alpha_F$ to account for 
hadronization and $\epsilon^2$ to account for final state dissociation. Amongst these, $\epsilon^2$ is already fixed from our previous 
analysis of $J/\psi$ production cross sections, as argued above. The other two parameters namely $\alpha_F$ and $f_{\psi(2S)}$ have to 
be fixed from the present data. This is in contrast to our earlier $J/\psi$ analysis where both $f_{J/\psi}$ and $\alpha_{F}$ were fixed 
from the analysis of inclusive $J/\psi$ production cross sections as a function of beam energy in $p+p$ collision. Unfortunately no 
such data are available for $\psi(2S)$ production, leaving us to determine all the free parameters from $p+A$ data only.
 
Fig.~\ref{fig1} shows the variation of the $\psi(2S)$ production cross section for different target nuclei, as a function of $<L>$, in 
$p+A$ collisions measured by NA50 Collaboration. The data are expressed as the ratio of the $\psi(2S)$ production cross 
sections in $p+A$ collisions to 
that in $p+Be$ collisions. One advantage of fitting this ratio is that we get rid of the multiplicative fitting parameter $f_{\psi(2S)}$. The two theory curves result from fitting the above data sets following two parameterisations of $F(q^2)$. The $<L>$ values as calculated in Glauber model framework for different target nuclei~\cite{NA50-450} and published with the data 
are used to generate the theoretical curves. The $\alpha_F$ values extracted from different data sets are given in Table~\ref{Tab1}. They are comparable within errors {\it i.e.} they show a very weak beam energy dependence. As we are fitting the data to two not very different beam energies (the corresponding $\sqrt{s_{NN}}$ are only 2 GeV apart), it is unlikely to expect a strong variation in the value of $\alpha_F$. However one might expect that it should not have a strong energy dependence given that it is related to the factorization scale $Q^2$ and not to the momentum fraction $x$. In fact while analyzing $J/\psi$ data in $p+N$ collisions, the corresponding $\alpha_F^{J/\psi}$ has been found to be constant over a very broad energy range. Unfortunately the paucity of data  prevents us to make such tests for $\psi(2S)$ production.  

We have also fitted the $\psi(2S)$-to-Drell-Yan ratio as shown in Fig.~\ref{fig2}.  In Drell-Yan (DY) process, a quark and an anti-quark from the nucleons of the two colliding nuclei annihilate to form a virtual photon which subsequently decays into a $\mu^{+}\mu^{-}$ pair. We have calculated the leading order DY cross section using the standard prescription~\cite{Vogt}. The only CNM effect incorporated is the nuclear modification of the quark distributions inside the target. The inclusive cross sections are obtained by integrating the double differential cross section within the suitable mass and rapidity range as appropriate for a particular data set. Like our previous analysis, the only free parameter in this case is $K_{eff}$ defined as $K_{eff}=f_{\psi(2S)}/K_{DY}$ where $f_{\psi(2S)} = K_{\psi(2S)} \times N_{\psi(2S)}$. $K_{DY}$ takes care of higher order effects in DY production. The $\alpha_F$ values are same as obtained from the fitting of the ratio $\sigma^{\psi(2S)}_{p+A}$ to $\sigma^{\psi(2S)}_{p+Be}$.

Finally we describe the double ratio for $\psi(2S)$-to-$J/\psi$ production cross sections in Fig.~\ref{fig3}. No parameter is tuned for this observable. The experimental data points are deduced from the measured values accounting for the appropriate propagation of errors. The theoretical curves result from the model analysis of the single ratios ($p+A$ to $p+Be$) for $J/\psi$ and $\psi(2S)$. The production cross section of $J/\psi$ or $\psi(2S)$  depends on the parameter $\alpha_F$ of the transition probability, $F_{c\bar{c} \rightarrow J/\psi / \psi(2S)}$.  As $\alpha_F$ is different the production cross sections are also different, leading to a decrease in the double ratio as a function of $<L>$. As expected both the curves give a satisfactory description of the data.
\begin{table}
\caption{Values of the parameter $\alpha_F$, for both Gaussian (G) and power-law (P) parametrizations, as fixed from $\psi(2S)$ production cross sections in $p+A$ collisions at SPS.}
\label{Tab1}
\begin{tabular}{|c|c|c|c|} \hline
 Data set & $\alpha_{F}^{(G)}$ (GeV) & $\alpha_{F}^{(P)}$
\\ \hline
NA50-400 & 1.07 $\pm$ 0.06 & 1.5 $\pm$ 0.3 
\\ \hline
NA50-450-HI & 1.09 $\pm$ 0.07 & 1.4 $\pm$ 0.3 
\\ \hline
NA50-450-LI & 1.20 $\pm$ 0.17  & 0.95 $\pm$ 0.54 
\\ \hline
\end{tabular}
\end{table}

\section{Predictions at FAIR energies}
\begin{figure} 
\includegraphics[height=5.5cm,width=5.5cm]{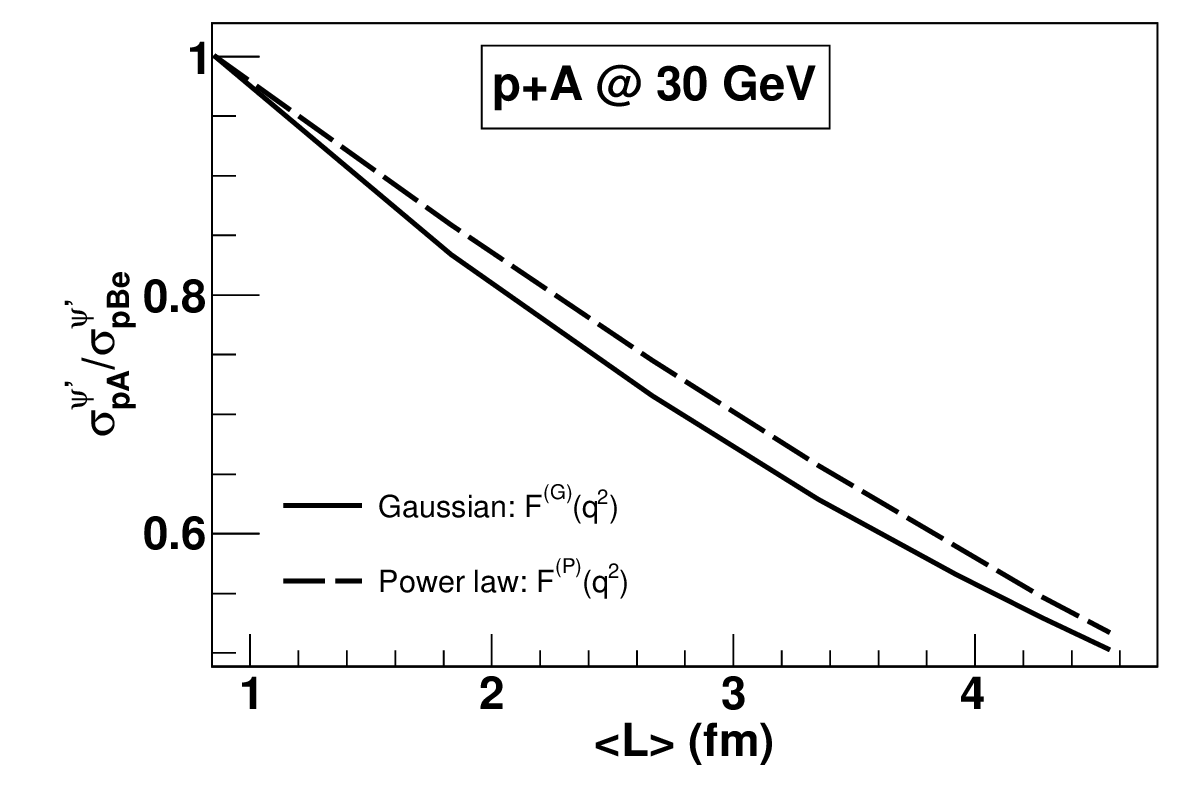}
\caption{\footnotesize Model prediction of $\psi(2S)$ production in 30 GeV $p+A$ collisions, within a rapidity interval $-0.5 \le y_{cms} \le 0.5$, at FAIR. Seven different target nuclei ($A= Be, Al, Cu, In, W, Pb, U$) are included in the calculation. Two theoretical curves represent two different mechanisms of resonance formation.}
\label{fig4}
\end{figure}
The goal of the Compressed Baryonic Matter (CBM) experiment at Facility for Antiproton and Ion Research (FAIR) is to explore the QCD phase diagram in the region of high net-baryon densities and moderate temperatures~\cite{CBM-physics}. This experiment is designed  to run at unprecedented high interaction rates which would enable precision measurements of rare diagnostic probes which are sensitive to the dense phase of the nuclear fireball. The foreseen program includes the measurement of charmonia ($J/\psi$, $\psi(2S)$) via their decay into di-leptons. 

The FAIR Modularised Start Version (MSV) comprises the SIS100 accelerator which would provide energies for gold beams up to 12 A GeV which is below the kinematic threshold for charmonium production ($E_{b}^{th} \simeq 12.2$ A GeV for $J/\psi$ and $E_{b}^{th} \simeq 15.6$ A GeV for $\psi(2S)$) and for protons up to 30 GeV. The research program of CBM at SIS100 includes the detailed measurements of charmonium production in $p+A$ collisions with varying target mass number. Such measurements are expected to shed light on the systematics of charmonium production at close to threshold energies and their interaction with cold nuclear matter. In addition, they constitute the necessary reference for future measurements at SIS300 accelerator, which will provide heavy ion beams up to 35 A GeV.

The basic motivation of this work is to understand mechanisms of $\psi(2S)$ production in proton induced collisions at SPS energies and then to extrapolate them to much lower energies at FAIR. Dynamics of charmonium production has been studied in 25 A GeV $Au+Au$ collisions using HSD transport model~\cite{hsd}. Two different mechanisms of anomalous suppression in nuclear collisions, namely the {\it hadronic co-mover scenario} and {\it QGP threshold scenario} have been investigated in detail. The centrality dependence of $\psi(2S)$ to $J/\psi$ ratio is found to be distinguishably different for the two cases with larger suppression for co-mover absorption. However a clear identification of anomalous $\psi(2S)$ suppression indeed demands for a correct estimate of the CNM dissociation effects. These can be correctly modelled in $p+A$ collisions, where production of any secondary medium at such low energies is usually not possible. 
   
\begin{figure} 
\includegraphics[height=4.cm,width=4.cm]{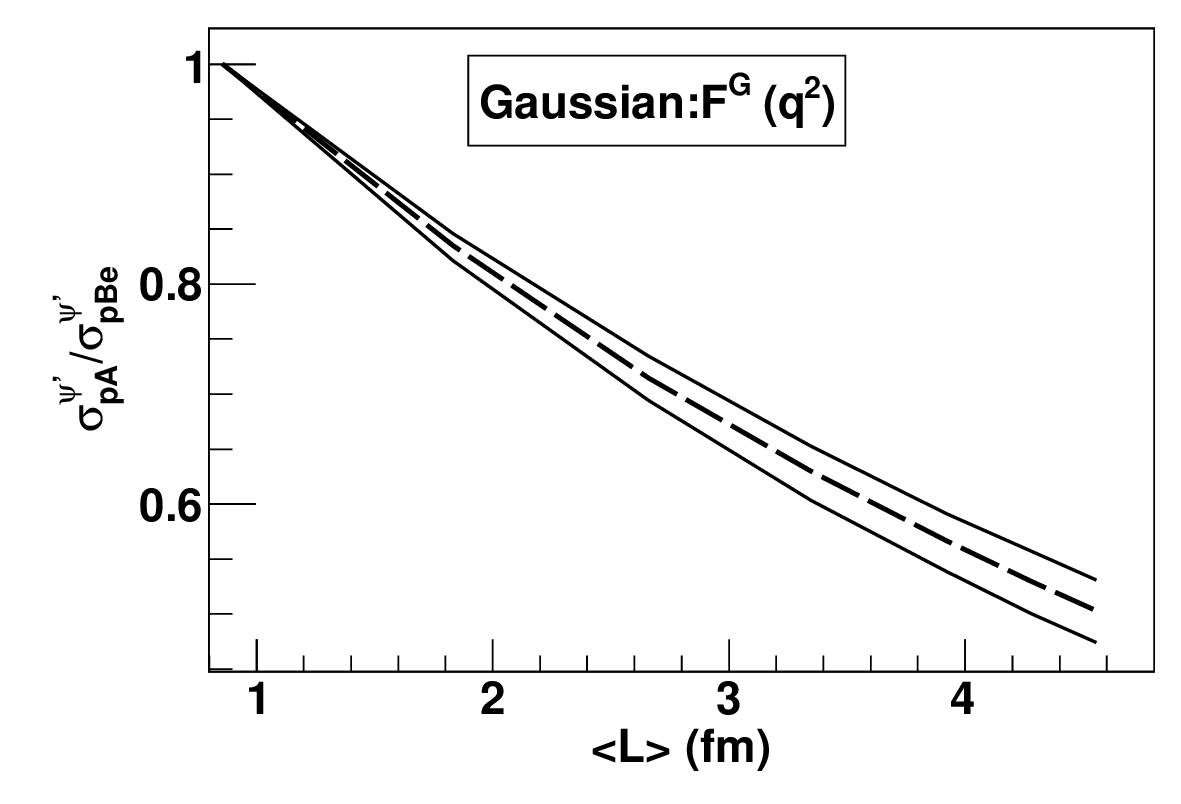}
\includegraphics[height=4.cm,width=4.cm]{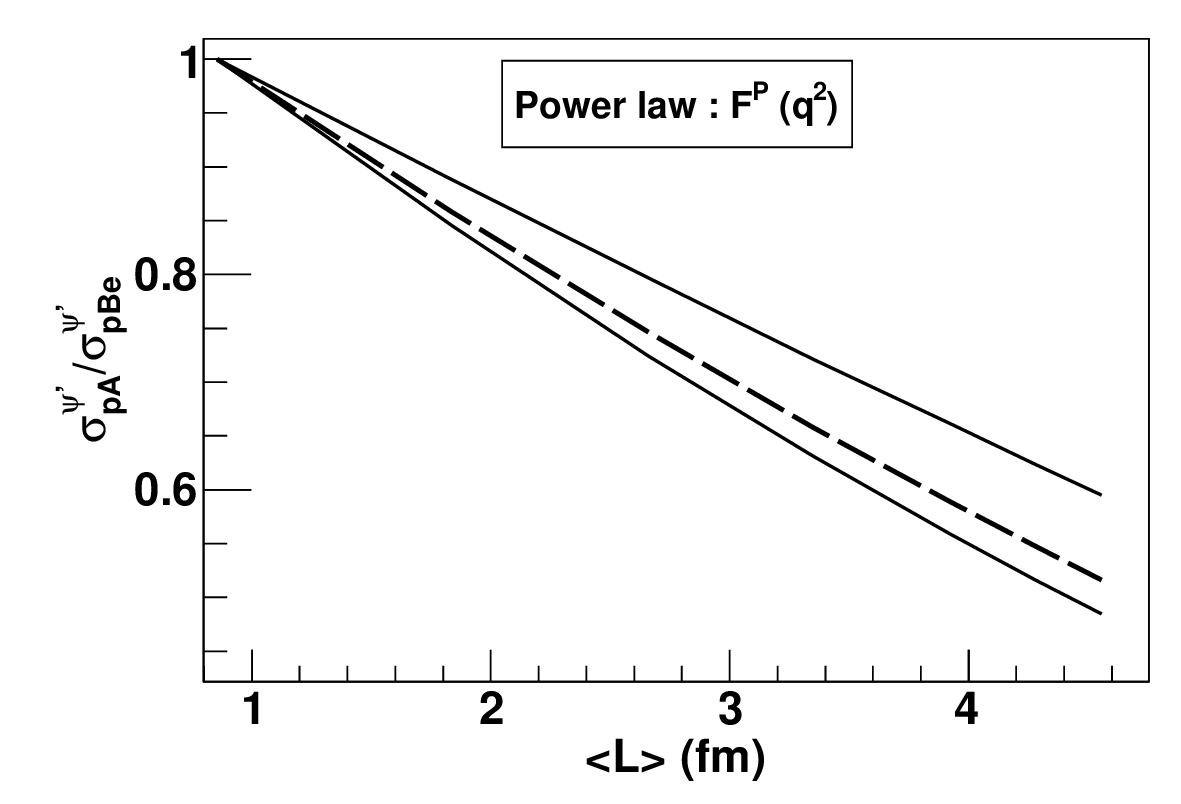}
\caption{\footnotesize Dependence of $\psi(2S)$ production on $\alpha_F$ in 30 GeV $p+A$ collisions, estimated in the kinematic domain $-0.5 \le y_{cms} \le 0.5$. Both Gaussian ($F^{G} (q^2)$) (left) and Power law ($F^{P}(q^2)$) (right) parameterisations of transition probability are included in our calculations. The bands represent the uncertainties 
in $\alpha_F (\alpha_F^{(G)} = 1.08 \pm 0.04$ and $\alpha_{F}^{(P)} = 1.4 \pm 0.2)$. For Gaussian case, the lower edge of the band corresponds to smaller value of $\alpha_F$ and vice versa for Power law.}
\label{fig5}
\end{figure}

We thus extrapolate our model calculations to estimate the $\psi(2S)$ production expected in 30 GeV $p+A$ collisions at FAIR SIS100 synchrotron. To do this we need to fix the values of $\alpha_F$ at relevant beam energies. As we saw in the previous section the $\alpha_F$ values obtained from fitting the SPS data are comparable within error bars. We have fitted those values to deduce beam energy independent constant values $\alpha_F^{(G)} = 1.08 \pm 0.04$ and $\alpha_{F}^{(P)} = 1.4 \pm 0.2$. 
These $\alpha_F$s are then used to estimate the cross sections at FAIR. The beam kinematic threshold for $\psi(2S)$ production in elementary collisions is $E_{b}^{th} \simeq 15.6$ GeV. Thus $\psi(2S)$ production may only be significant at top SIS100 energy. 
Fig.~\ref{fig4} represents the $\psi(2S)$
production cross section in 30 GeV $p+A$ collisions for seven different nuclear targets ($A= Be, Al, Cu, In, W, Pb, U$). Results are expressed in terms of the ratio of inclusive production cross sections, evaluated over a rapidity slice $-0.5 \le y_{cms} \le 0.5$. In ref.~\cite{partha1} the $\epsilon^2$ values were obtained from fitting the data of different fixed target experiments with different energy of the incident proton beam, in the range 158 - 920 GeV. The extracted best fit values of  $\epsilon^2$ were found to be sensitive to the employed form of transition probability and the underlying PDF set (different values of $\epsilon^2$ for free proton PDF and EPS09 nPDF). Both $\epsilon^2_G$ and $\epsilon^2_P$ show a non-negligible energy dependence with $\epsilon^2_{P(G)}$ increasing with decrease in the collision energy. To derive the level of suppression at lower energies relevant for FAIR, the observed dependences of $\epsilon^2$, extracted for EPS09 nPDF set, on the energy of the incident proton beam have been parameterised, for both
$F^{(G)}(q^2)$ and $F^{(P)}(q^2)$ using exponential functions. This gives  corresponding values at 30 GeV as $\epsilon^{2}_{(P)} = 0.31 \pm 0.05$ and $\epsilon^{2}_{(G)} = 0.44 \pm 0.08$. The amount of suppression is larger than that measured at SPS. As investigated in ref.~\cite{partha3}, this is caused by convolution of two CNM effects, the shadowing of the parton densities inside the target nuclei and the amplification of final state dissociation in the kinematic region probed at FAIR. Note that in our model both the CNM effects are same for both $J/\psi$ and $\psi(2S)$. However unlike SPS, at FAIR the model results indicate a relatively larger suppression with Gaussian parameterisation ($F^{(G)} (q^2)$) compared to power law ($F^{(P)} (q^2)$). This can be attributed to the non-uniform phase space distribution of the pre-resonant $c\bar{c}$ pairs, an effect that gets amplified with decrease in beam energy. It may also be interesting to compare fitted values of $\alpha_F$ to the previously obtained $J/\psi$ results. For $F^{(G)} (q^2)$, $\alpha_F$ is the width of the Gaussian probability distribution for a $c\bar{c}$ pair to form a resonance and the corresponding $\alpha_{F}^{J/\psi}$ is larger than $\alpha_{F}^{\psi(2S)}$. The $J/\psi$ being significantly below the open charm threshold it would naturally have a larger production cross section. For power law case, $\alpha_F^{(P)}$ is related to the probability of soft gluon radiation, required for 
color neutralisation. As evident from Eq.~\ref{power}, a smaller $\alpha_F$ leads to increased resonance production cross section.   

\begin{figure} 
\includegraphics[height=4.cm,width=4.cm]{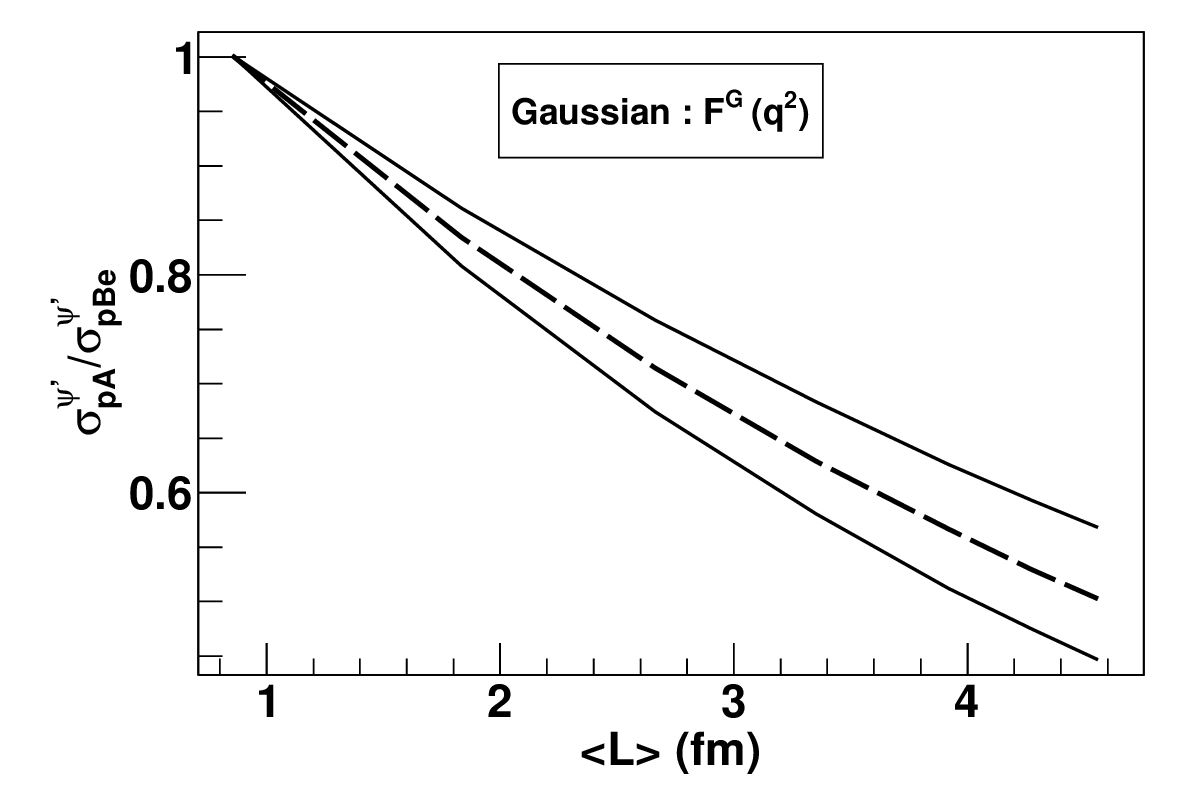}
\includegraphics[height=4.cm,width=4.cm]{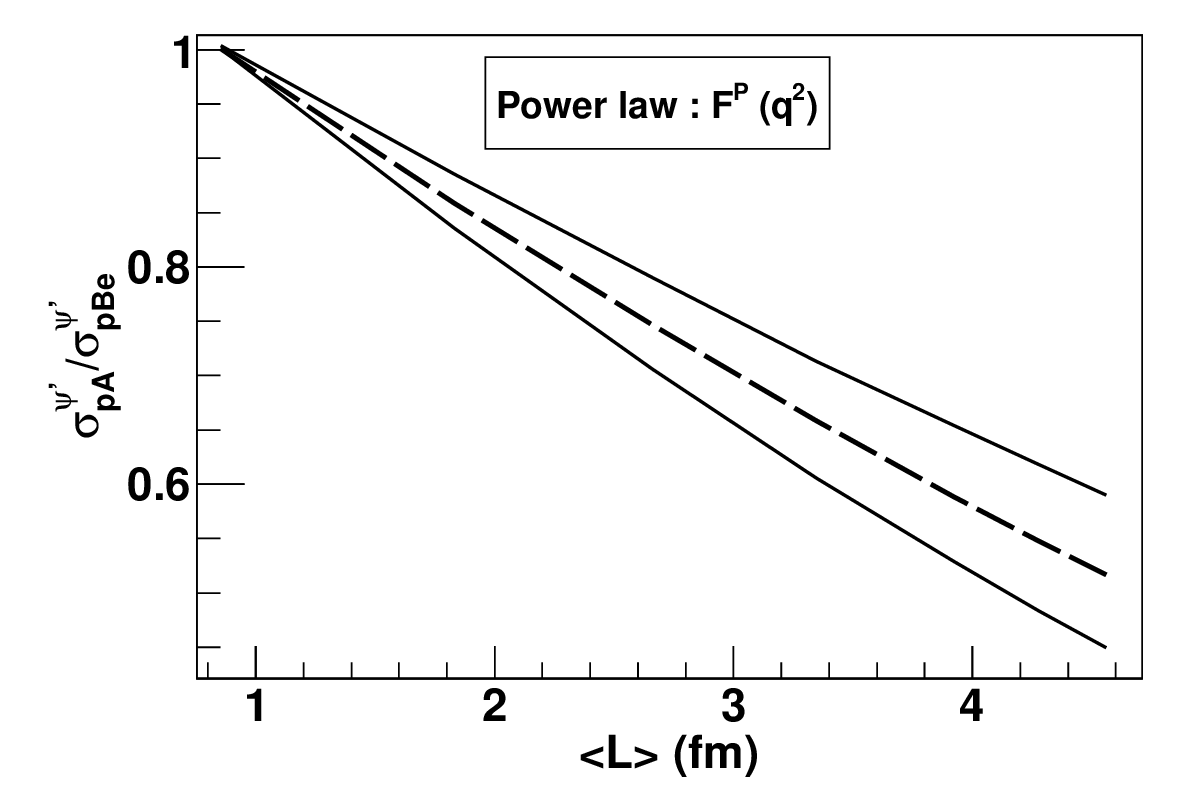}
\caption{\footnotesize Dependence of $\psi(2S)$ production on $\epsilon^2$, within the rapidity domain $-0.5 \le y_{cms} \le 0.5$, in 30 GeV $p+A$ collisions. Both Gaussian ($F^{G} (q^2)$) (left) and Power law ($F^{P}(q^2)$) (right) parameterisations of transition probability are included in our calculations. The bands represent the uncertainties in $\epsilon^{2}$ ($\epsilon^{2}_{(P)} = 0.31 \pm 0.05$ and $\epsilon^{2}_{(G)} = 0.44 \pm 0.08$). For both the cases lower edge of the band correspond to the larger value of $\epsilon^2$ signifying stronger suppression.} 
\label{fig6}
\end{figure}

It is also interesting to test the sensitivity of the model predictions to the uncertainties in the input parameters of the model. The variation of $\psi(2S)$ production cross sections for different target nuclei, with  $\alpha_F$ is shown in Fig.~\ref{fig5}. As expected, for Gaussian case, the cross section increases with increase in $\alpha_{F}^{G}$. For the power law case, larger value of $\alpha_{F}^{P}$ leads to reduced resonance formation. The asymmetry of $F^{(P)}(q^2)$ in $\alpha_F$ results in an asymmetric cross section 
ratio. The dependence of $\psi(2S)$ production on $\epsilon^{2}$ is shown in Fig.~\ref{fig6}. For both the parametrisation of formation probability, increment in $\epsilon^2$ leads to larger suppression. Finally in our calculations, we set $m_C=1.5$ GeV and $m_D = 1.85$ GeV. Variation of $m_C$ ($\Delta m_{C} = \pm 0.2$ GeV) and $m_D$ (within the vacuum masses of different members of $D$-family ($D^\pm$ and $D^{0}$)), slightly ($< 1 \%$) changes the resonance production cross sections.

For Gaussian parametrisation, in the spirit so called ``$\rho$L parametrisation'' of Glauber model~\cite{NA50-400}, we can also find an absorption cross section~\cite{partha1} for $\psi(2S)$ as $\sigma_{abs}^{\psi(2S)} = 12.4 \pm 2.2$ mb. The corresponding value for $J/\psi$ meson was found to be $\sigma_{abs}^{J/\psi} = 10.1 \pm 1.77$ mb.

In fact being close to the kinematic threshold, the $\psi(2S)$ production cross section will also be extremely low at FAIR. In ref.~\cite{hsd}, the authors have provided empirical formula (parametrisation) to obtain inclusive $\psi(2S)$ production cross section as a function of $\sqrt{s}$, in elementary $p+N$ collisions. At a beam energy of 30 GeV, this corresponds to an inclusive production cross section of $\sigma^{NN}_{\psi(2S)} \simeq 0.1$ nb. Following the so called $\alpha$ parametrisation~\cite{Satz}, the corresponding production cross section for a nuclear target of mass number $A$ is given by, $\sigma^{pA}_{\psi(2S)} = \sigma^{NN}_{\psi(2S)} \times A^{\alpha}$. For a typical value of $\alpha=0.95$ the corresponding inclusive $\psi(2S)$ production cross section in 30 GeV $p+Au$ collisions comes to be 15 nb.  The related $\psi(2S)$ yield in di-muon  channel will be very low, making their detection extremely challenging. To enable the measurement of such rare probes, the FAIR accelerators are being designed with maximum foreseen beam intensity as high as $3 \times 10^{13}/$s for protons and $10^{9}/$s for heavy ions~\cite{peter}, and detectors with extremely high rate capabilities~\cite{much-tdr}. It may be noted here that in addition to perturbative production, new mechanisms of charmonium production are proposed in literature~\cite{steinheimer,kiselev} at near threshold beam energies. Sub-threshold production of charmonia 
via decay of massive baryonic resonances produced by multi-step collision of nucleons~\cite{steinheimer} might increase the $\psi(2S)$ yield in low energy collisions and thereby facilitate their detection.  It has also been shown~\cite{steinheimer1} that Fermi momenta of the nuclei play only a minor role  in particle production. In this context, the reader should also take note of the fact that the $p+A$ data at 30 GeV can not be directly used to estimate the CNM effects in $\psi(2S)$ production in nuclear collisions at sub-threshold regime.

Even though we explicitly talk about FAIR, our estimations will be useful for any other existing or future facility like NICA~\cite{nica} or J-PARC~\cite{j-parc} which would aim for investigation of CNM effects in charmonium production at near threshold energies. Of course the essential requirement to make such measurements feasible is to have proton beams with very high intensity and very fast detectors and electronics to cope with resulting high interaction rates.

Limitations on the applicability of our model calculations at FAIR energy domain may be noted. QVZ model is based on QCD factorization and perturbative production of the $c\bar{c}$ pairs, which in general is operative at high energies. Here we assume that factorization still holds for charmonium production close to threshold, which is not free from doubt. An alternative approach based on scaling arguments for near threshold quarkonium production can be found in ref.~\cite{partha4}. 
Moreover the model assumes propagation of the perturbatively produced $c\bar{c}$ pairs through the nucleus and the non-perturbative transition from $c\bar{c} \rightarrow \psi$ (2S) occurs outside the nucleus. The validity of this assumption is also questionable at low energies like those at SPS and FAIR owing to smaller Lorentz dilation of the intrinsic resonance formation times in the target rest (laboratory) frame. However determination of the quarkonium formation times are far from being unique and strongly model dependent~\cite{KS,KZ,HeHK,KT,BO}. At a beam energy of 158 GeV, with an intrinsic formation time of $\tau_0 \simeq 0.3$ fm~\cite{HeHK}, the J/$\psi$ formation length in the laboratory frame ($l_{J/\psi}$), at mid-rapidity (corresponding to $x_F=0$) is around 3 fm. Even then QVZ model can describe all the available data for J/$\psi$ production in both 400 GeV and 158 GeV $p+A$ collisions collected by the NA50 and NA60 Collaborations at SPS~\cite{partha1}. The model also gives a satisfactory description of the latest data on J/$\psi$ suppression measured in nuclear collisions for both 158 A GeV $In+In$ and $Pb+Pb$ collisions as measured by NA60 and NA50 Collaborations respectively~\cite{partha2}. For $\psi(2S)$ production let us take the intrinsic formation time as $\tau_0 \simeq 0.91$ fm~\cite{KT}. At SPS for $E_b=400$ GeV this would correspond to a $x_F$ range $-0.12 < x_F < 0.16$ and formation length ($l_{\psi(2S)}$) range $8.6 < l_{\psi(2S)} < 22.8$ fm, for the measured di-muon rapidity interval $-0.425 < y_{cm} < 0.575$. Hence the assumption of resonance formation taking place outside the nucleus in $p+A$ collisions is satisfied. As described in the present manuscript, QVZ model gives a reasonable description of the data. Unfortunately there are no data on $\psi(2S)$ production in 158 GeV $p+A$ collisions to test the model calculations. At FAIR for $E_b=30$ GeV, a rapidity interval of $-0.5 < y_{cm} < 0.5$ corresponds to almost same $x_F$ range ($-0.498 < x_F < 0.498$). The corresponding $l_{\psi(2S)}$ (with $\tau_0^{\psi(2S)} = 0.91$ fm) ranges from 2.5 fm to 5.2 fm (Note that for $\tau_0^{\psi(2S)} = 1.5$ fm~\cite{KS}, $l_{\psi(2S)}$ will be accordingly larger). Charmonium data to be collected at FAIR SIS100 are thus highly welcome to validate or nullify some of the above arguments.   
In this work the model includes two most important CNM effects namely initial state shadowing/anti-shadowing and final state dissociation of the $c\bar{c}$ pairs and has found to describe all the available data set on $J/\psi$ and $\psi'$ production in $p+A$ collisions at SPS energy domain. This certainly makes it viable for making predictions at FAIR energies, which can be tested as soon as we collect data from SIS100.

\section{Summary}

In summary, we have analysed the available SPS data on $\psi(2S)$ production in $p+A$ collisions measured by NA50 Collaboration. The adapted version of QVZ model is employed for this purpose. The model has been found earlier to give reasonable description of $J/\psi$ production at SPS. Non-availability of suitable data at low energy collisions, makes the analysis rather complex by adding large uncertainties in determination of the model parameters. For both $\psi(2S)$ and DY initial state modification of the parton distributions inside the target are taken into account. Within the QVZ approach, the final state dissociation of the $\psi(2S)$ mesons inside the cold nuclear medium is accounted through the multiple scattering of the pre-resonant $c\bar{c}$ pairs with the spectator nucleons. The model is found to describe the observed suppression for both forms of the transition probability. Model calculations are extrapolated to the FAIR energy domain. A much larger suppression is expected at SIS100 $p+A$ collisions. Data from SIS100 should be able to remove large uncertainties in the parameters and thus can lead to a much precise estimate. With such precise estimates in $p+A$ scenario one can plan to extrapolate to $A+A$ scenario once SIS300 synchrotron ring 
becomes operative. 

\section{Acknowledgement} We thank Saumen Datta for useful discussions. AB thanks UGC \& BRNS for support. We are also thankful
to Volker Friese for fruitful discussions.

\end{document}